\documentclass[aps,prl,showpacs, letterpaper,twocolumn,superscriptaddress]{revtex4-1}

\usepackage{amsmath, amsfonts}
\usepackage{subfig, natbib}
\usepackage{graphicx,color}
\usepackage{multirow}
\usepackage{algorithm}

\newcommand{\ud}{\,\mathrm{d}}

\newcommand{\norm}[1]{\lVert#1\rVert}

\newcommand{\wt}[1]{\widetilde{#1}}

\newcommand{\CS}{\mathcal{C}}

\newcommand{\eg}{\textit{e.g.}}
\newcommand{\Or}{\mathcal{O}}

\setcitestyle{super}

\begin{document}

\title{Compressed representation of Kohn-Sham orbitals \\ via selected
columns of the density matrix}

\author{Anil Damle}
\email[Corresponding author: ]{damle@stanford.edu}
\affiliation{Institute for Computational and Mathematical Engineering,
Stanford University, Stanford, CA 94305}

\author{Lin Lin}
\affiliation{Department of Mathematics, University of California,
Berkeley, Berkeley, CA 94720}
\affiliation{Computational Research Division, Lawrence Berkeley National Laboratory, Berkeley, CA 94720}

\author{Lexing Ying}
\affiliation{Department of Mathematics, Stanford University, Stanford, CA 94305}
\affiliation{Institute for Computational and Mathematical Engineering,
Stanford University, Stanford, CA 94305}

\begin{abstract}
  Given a set of Kohn-Sham orbitals from an insulating system, we
  present a simple, robust, efficient and highly parallelizable method
  to construct a set of, optionally orthogonal, localized basis functions for the associated
  subspace.  Our method explicitly uses the fact that density matrices 
  associated with insulating systems decay exponentially along the off-diagonal
  direction in the real space representation.
  Our method avoids the usage of an optimization procedure,
  and the localized basis functions are constructed directly from a set of
  selected columns of the density matrix (SCDM). 
  Consequently, the only adjustable parameter in our method is the truncation
  threshold of the localized basis functions.
  Our method can be used in any
  electronic structure software package with an arbitrary basis set.  
  We demonstrate the numerical accuracy and parallel scalability of the
  SCDM procedure using orbitals generated by the Quantum ESPRESSO
  software package.  We also demonstrate a procedure for combining SCDM
  with Hockney's algorithm to efficiently perform Hartree-Fock exchange
  energy calculations with near linear scaling.  

\end{abstract}

\maketitle

%%%%%%%%%%%%%%%%%%%%%%%%%%%%%%%%%%%%%%%%%%%%%%%%%%%%%%%%%%%%%%%%%%%%%%%%%%%%%%%
\section{Introduction}\label{sec:intro}
%%%%%%%%%%%%%%%%%%%%%%%%%%%%%%%%%%%%%%%%%%%%%%%%%%%%%%%%%%%%%%%%%%%%%%%%%%%%%%%

Kohn-Sham density functional theory
(KSDFT)~\cite{HohenbergKohn1964,KohnSham1965} is the most widely used
electronic structure theory for molecules and systems in condensed
phase. In KSDFT the many body electronic structure properties are, in
principle, exactly mapped into a fictitious single particle system.  The
Kohn-Sham orbitals, which are orthonormal eigenfunctions of the single
particle Kohn-Sham Hamiltonian operator, can describe various physical
quantities such as density, energy, and atomic forces.  However, it is
expensive to compute and store the Kohn-Sham orbitals of large systems,
which are spatially delocalized. 
Let $N$ be the
number of degrees of freedom and $N_{e}$ be the number of electrons in
the system. The cost for storing the Kohn-Sham
orbitals is $\Or(N N_{e}),$ and the cost for computing them is generally
$\Or(N N_{e}^2),$ or ``cubic scaling'' assuming $N\sim \Or(N_{e})$.  In
modern KSDFT calculations the Hartree-Fock exact exchange term is also
often taken into account in the form of hybrid
functionals~\cite{PerdewErnzerhofBurke1996,Becke1993}.  The
computational cost for this step not only scales cubicly but also has a
large pre-constant, which limits the application of hybrid functional
calculations to hundreds of atoms.

In order to reduce both the storage requirements and the cost associated
with subsequent computations, it is important to realize that the
Kohn-Sham orbitals are not unique. Any non-degenerate linear
transformation of the set of Kohn-Sham orbitals yields exactly the same
physical properties of a system. In other words, the physically relevant
quantity is the subspace spanned by the Kohn-Sham orbitals.  Various
efforts~\cite{MarzariVanderbilt1997,WannierReview,Gygi2009,ELiLu2010,OzolinsLaiCaflischEtAl2013}
have been made to utilize this fact and to find a set of localized
orbitals to form compressed representation of a Kohn-Sham subspace.  For
example, the
Marzari-Vanderbilt~\cite{MarzariVanderbilt1997,WannierReview}
construction of maximally localized Wannier functions (MLWFs) uses a
nonlinear optimization approach to find a unitary transformation of the
Kohn-Sham subspace into a set of orthogonal functions localized in
real space.  The locality, or the ``nearsightedness''
principle, is guaranteed for insulating systems with a finite HOMO-LUMO
gap~\cite{Kohn1996,ProdanKohn2005}. The localized orbitals have wide
applications in chemistry and physics.  For instance, localized orbitals
can be be used to construct linear scaling methods for solving KSDFT
using local and semi-local
functionals~\cite{Goedecker1999,BowlerMiyazaki2012}, and for
calculations using hybrid
functionals~\cite{WuSelloniCar2009,GygiDuchemin2012}.

In this manuscript, we present an alternative method to the widely used
MLWFs to compute a set of localized orbitals associated with the Kohn-Sham
subspace for insulating systems. Our method is simple, robust, efficient
and highly parallelizable.  Our method explicitly uses the fact that for
insulating systems the single particle density matrix is exponentially
localized along the off-diagonal direction in the real space
representation~\cite{BenziBoitoRazouk2013,Kohn1996,Blount,Cloizeaux1964a,Cloizeaux1964b,
Nenciu}.  Our algorithm finds a set of localized orbitals by directly
using selected columns of the density matrix (SCDM) associated with the
Kohn-Sham orbitals. Consequently, by construction the SCDM are localized in the
real space representation.  
As opposed to the MWLFs which are orthonormal, the SCDM are not orthonormal in
general. However, a set of orthonormal and localized functions spanning the
Kohn-Sham subspace can be obtained via a simple linear transformation of
the SCDM.  

The only adjustable parameter in our method is the truncation threshold
for the localized functions, and our method can be
used in any electronic structure software package with any basis set,
ranging from planewaves to Gaussian basis sets, provided that the
Kohn-Sham orbitals can be represented on a real space grid.
One striking feature of our method is its
simplicity: a prototype sequential implementation takes just a few lines
of code. The construction of the SCDM only involves simple linear
algebra routines such as a rank-revealing QR factorization and
matrix-matrix multiplication. Therefore, the parallel implementation for
computing the SCDM can straightforwardly scale, for the problem size we tested, to more than one thousand
processors. This enables the computation of localized basis functions
for the self-consistent treatment of the Hartree-Fock terms in KSDFT
calculations with hybrid exchange-correlation functionals. As an application, we also demonstrate a procedure for
combining SCDM with Hockney's algorithm to efficiently compute the 
Hartree-Fock exchange energy.

%%%%%%%%%%%%%%%%%%%%%%%%%%%%%%%%%%%%%%%%%%%%%%%%%%%%%%%%%%%%%%%%%%%%%%%%%%%%%%%
\section{Theory}\label{sec:localization}
%%%%%%%%%%%%%%%%%%%%%%%%%%%%%%%%%%%%%%%%%%%%%%%%%%%%%%%%%%%%%%%%%%%%%%%%%%%%%%%

For insulating systems,  the locality of the single
particle density matrix along the off-diagonal direction can be in
general
observed in the real space
representation~\cite{Kohn1996,ProdanKohn2005}. For the sake of clarity, in this
manuscript we explicitly require the Kohn-Sham orbitals to be
represented on a real space grid defined as below.  In some cases the
real space representation may not be necessary, and we postpone such a discussion to the conclusion.

Let $\{\psi_{j}(x)\}_{j=1}^{N_{e}}$ be a set of Kohn-Sham
orbitals which satisfy the orthonormality condition
\begin{equation}
  \int \psi_{j}^*(x) \psi_{j'}(x) \ud x = \delta_{jj'},
  \label{eqn:orthonormal}
\end{equation}
and we have access to $\psi_{j}(x)$ evaluated at a set of discrete grid
points $\{x_{i}\}_{i=1}^{N}$.  Let $\{\omega_{i}\}_{i=1}^{N}$ be a set
of positive integration weights associated with the grid points
$\{x_{i}\}_{i=1}^{N}$, then the discrete orthonormality condition
corresponding to Eq.~\eqref{eqn:orthonormal} is given by
\begin{equation}
  \sum_{i=1}^{N} \psi_{j}(x_{i}) \psi_{j'}(x_{i}) \omega_{i} =
  \delta_{jj'}.
  \label{eqn:orthonormal_discrete}
\end{equation}
Let $\psi_{j}=[\psi_{j}(x_{1}), \psi_{j}(x_{2}), \ldots,
\psi_{j}(x_{N})]^{T}$ be a column vector, and $\wt{\Psi}=[\psi_{1},  \ldots,  \psi_{N_e}]$
be a matrix of size $N\times N_{e}$. In this paper we call $\wt{\Psi}$ the
\textit{real space representation} of the Kohn-Sham orbitals.  We
also define a diagonal matrix
$W=\mathrm{diag}[\omega_{1},\ldots,\omega_{N}]$.

It should be noted that the real space representation of Kohn-Sham
orbitals can be obtained with any type of basis set, and therefore our
method is applicable for any electronic structure software package. For instance, if
the Kohn-Sham orbitals are represented using the planewave basis
functions, their real space representation can be obtained on a uniform
grid efficiently with the fast Fourier transform (FFT) technique. In
such case, $\omega_{i}$'s take the same constant value for all $i$.  For
general basis sets such as Gaussian basis functions or localized atomic
orbitals, let $\{\chi_{k}(x)\}_{k=1}^{M}$ be the collection of basis
functions (usually $M\ll N$). We first evaluate each basis function on
the real space grid as $\chi_{k}=[\chi_{k}(x_{1}), \chi_{k}(x_{2}),
\ldots, \chi_{k}(x_{N})]^{T}$, and denote by
$X=[\chi_{1},\ldots,\chi_{M}]$ the collection of all basis functions. 
$X$ is a matrix of size $N\times M$.  Then the Kohn-Sham orbitals can be
obtained as the linear combination of basis functions as $\wt{\Psi} = X
Z$.  Here the matrix $Z$ of size $M\times N_{e}$, and is usually
obtained by solving a generalized eigenvalue problem.  For these general
basis functions and in particular in all-electron calculations, the grid
points are usually chosen to be non-uniform to improve the accuracy of
numerical quadrature, and correspondingly the weights $\omega_{i}$'s
are non-uniform as well.

We define $\Psi=W^{\frac12} \wt{\Psi}$ to be the set of weighted Kohn-Sham
orbitals represented in the real space, in which case the discrete orthonormality
condition in Eq.~\eqref{eqn:orthonormal_discrete} becomes
$\Psi^{*}\Psi=I$, where $I$ is an identity matrix of size $N_{e}$.
We now seek a compressed basis for the span of $\Psi$, denoted by the
set of vectors $\Phi=[\phi_{1}, \ldots,  \phi_{N_e}]$, such that each
$\phi_{i}$ is localized. 
The single particle density matrix is defined as
$P=\Psi \Psi^{*}$. The nearsightedness principle states that  for
insulating systems, each column of the matrix $P$ is localized. As a result, selecting any linearly independent subset of $N_e$ of them
will yield a localized basis for the span of $\Psi.$ However, picking
$N_e$ random columns of $P$ may result in a poorly conditioned basis if,
for example, there is too much overlap between the selected
columns.  Therefore, we would like a means for choosing a well
conditioned set of columns, denoted $\CS,$ to use as the localized
basis. Intuitively we expect such a basis to select columns to minimize
overlaps with each other when possible.

To achieve this we utilize the so called \textit{interpolative decomposition} \cite{cheng2005compression}. Given an $N\times N$ matrix $A$ with rank $k$, such a factorization seeks
to find a permutation matrix $\Pi$, a subset of $k$ columns of $A$ whose
indices form the set $\CS$ and a matrix $T$ such that
\begin{equation}
A\Pi = A_{:,\CS} \begin{bmatrix} I & T \end{bmatrix},
\end{equation}
and $\norm{T}$ should be small.
The interpolative decomposition can be computed via the (strong) rank
revealing QR factorization~\cite{cheng2005compression,gu1996efficient}. Such a factorization computes
${A\Pi = Q \begin{bmatrix}R_{1} & R_{2}\end{bmatrix}}$ where $Q$ is an $N\times k$ orthonormal
matrix, $R_{1}$ is an upper triangular matrix and $\Pi$ is a permutation
matrix. The permutation $\Pi$ is chosen to keep $R_{1}$ well
conditioned. The interpolative decomposition is a powerful technique,
and can be used to build low rank approximations for a matrix $A$ with
many small singular values (i.e. $A$ has many columns that are nearly
linearly dependent). In such cases, algorithms for constructing
these factorizations often choose $k$ to ensure that a certain relative
accuracy in the approximation of $A$ is achieved. However, in our
situation we know $P$ is exactly of rank $N_e$ and are thus not concerned
with approximation error.

It may not be feasible to compute an interpolative decomposition of $P$
directly because we would have to construct $P$ explicitly. The
storage cost of $P$ is $N^{2}$, and
computing a partial rank revealing QR factorization of $P$ 
scales as $\Or(N_e N^2),$ which is prohibitively expensive. Randomized algorithms exist that would reduce this cost, for example \cite{martinsson2011randomized,liberty2007randomized}, but they do not achieve the desired computational scaling. To achieve the desired scaling, observe that
for any rank revealing QR factorization of $\Psi^*$
\[
\Psi^*\Pi=QR
\]
where $Q$ is an $N_e\times N_{e}$ matrix with orthonormal columns, then 
\[
P \Pi = (\Psi Q) R.
\]
It can be readily verified that $\Psi Q$ is an $N \times N_{e}$
matrix with orthonormal columns. Therefore, rather than computing a rank revealing QR factorization of $P$ we may equivalently compute a rank revealing QR factorization of $\Psi^{*}.$ This computation scales as
$\Or(N_{e}^2 N).$ The permutation matrix $\Pi$ reveals the selection
of columns $\CS,$ and the SCDM can be computed as
\[
\tilde{\Phi}\equiv P_{:,\CS}=\Psi \Psi_{\CS,:}^{*}.
\]

Using the SCDM we may recover the density matrix in a simple manner. Since $P$ is a
rank $N_{e}$, Hermitian and positive semi-definite matrix and $P_{:,\CS}$ has linearly independent columns we may write $P$ in the
form $P = P_{:,\CS} D P_{:,\CS}^{*}$, where $D$ is an $N_{e}\times
N_{e}$ matrix. By restricting $P$ to the $\CS$ row and column
indices, $P_{\CS,\CS} = P_{\CS,\CS} D P_{\CS,\CS}^{*}$ we observe that
$D=\left(P_{\CS,\CS}\right)^{-1}$ is uniquely determined. Therefore,
\begin{equation}
P = P_{:,\CS} \left(P_{\CS,\CS}\right)^{-1} P_{:,\CS}^{*}.
  \label{eqn:dmformula}
\end{equation}

Eq.~\eqref{eqn:dmformula} also suggests a method for the construction of orthonormal and
localized basis functions.  If we let 
\[
P_{\CS,\CS} = LL^*
\]
be a Cholesky factorization of $P_{\CS,\CS}$, then 
\[
\Phi = \tilde{\Phi} L^{-*}
\]
has orthogonal columns and they may be used as a basis for the
span of $\Psi.$ In this case we may also write $P = \Phi\Phi^{*}$. The orthogonality of
$\Phi$ follows from $P$ being an orthogonal projector.
Based on the locality of the columns of $P$ and because the permutation
$\Pi$ picks columns of $P$ that form a well conditioned basis,
$P_{\CS,\CS}$ is sparse and only lower triangular entries of $L$ near the diagonal have significant magnitude. Therefore, this orthogonalization step does not significantly
impact the localization of the basis functions.

Concisely, we construct the SCDM $\tilde{\Phi}$ or the
orthogonalized SCDM $\Phi$ via the algorithm below.
\begin{enumerate}
\item Compute the index set $\CS$ associated with an interpolative
  decomposition of $\Psi^*$ via a rank revealing QR decomposition.
\item Compute the SCDM $\tilde{\Phi} = P_{:,\CS} = \Psi (\Psi_{\CS,:})^{*} $ as the new localized basis

\noindent If the orthogonalized SCDM are desired, then

\item Compute the Cholesky factorization $ P_{\CS,\CS} = LL^*$
\item Compute the orthogonal basis $\Phi$ by solving ${\Phi L^* = \tilde{\Phi}}$  
\end{enumerate}
This algorithm is completely deterministic and can be applied to any
local or nonlocal basis set. In general, when computing the rank
revealing QR factorization for an interpolative decomposition, a so
called strong rank revealing QR factorization is technically required.
However, for the types of systems we are interested in a more
traditional rank revealing QR factorization, such as the one implemented
in LAPACK \cite{lapack} as DGEQP3 suffices. The use of a Cholesky
factorization allows us to avoid explicit inversion of $P_{\CS,\CS}$ and
instead use triangular solves to either apply the spectral projector or
orthogonalize the basis.

The overall cost of the algorithm  is $O(N_e^2 N + N_e^3)$ to build
$\Phi$ or $\tilde{\Phi}.$ The cost of the necessary rank revealing QR is
$\Or\left(N_e^2 N\right)$ and the dominant cost in $N_e$ that has no dependence on $N$ is the Cholesky
factorization, which costs $\Or(N_e^3)$. The cost of forming the columns
of $\tilde{\Phi}$ is $\Or\left(N_e^2 N\right)$ as is the cost of
constructing $\Phi$ via triangular solves.  Because the algorithm is
constructed from simple and standard linear algebra routines it is easy
to parallelize. Specifically, steps one, two and four in the preceding
algorithm may each use a parallel version, \eg~from ScaLAPACK
\cite{Scalapack}, of the respective linear algebra routine. Because the
required Cholesky factorization is of a $N_e \times N_e$ matrix it is
often not necessary to parallelize that portion of the algorithm, though
one certainly could. By utilizing common factorizations and operations
the algorithm immediately benefits from improvements to the underling,
serial or parallel, linear algebra routines and does not require any
specialized code. One example of this would be use of recent variants on
computing a parallel rank revealing QR factorizations such as the
communication avoiding version discussed in \cite{DemmelRRQR}. The
efficiency of a parallel version will be briefly discussed in the
section on numerical results.

To illustrate the simplicity of the algorithm we present a serial
implementation of the algorithm in MATLAB \cite{MATLAB}. The following
two lines of code implement the above algorithm and compute the
SCDM. The discrete input $\Psi$
is represented as \verb+Psi+ and \verb+Ne+ is the column dimension of
$\Psi$, and therefore the number of localized basis functions. When
complete, the matrix \verb+Pc+ contains the localized SCDM. 
\begin{verbatim}
[Q, R, piv] = qr(Psi',0);
Pc = Psi*(Psi(piv(1:Ne),:)');
\end{verbatim}
If a set of orthonormal and localized orbitals are to be computed, the
following three lines can be added, and the matrix \verb+Phi+ contains
the orthogonalized and localized SCDM.
\begin{verbatim}
S = Pc(piv(1:Ne),:);
Rchol = chol(S);
Phi = Pc/(Rchol);
\end{verbatim}

%%%%%%%%%%%%%%%%%%%%%%%%%%%%%%%%%%%%%%%%%%%%%%%%%%%%%%%%%%%%%%%%%%%%%%%%%%%%%%%
\section{Numerical results}\label{sec:app}
%%%%%%%%%%%%%%%%%%%%%%%%%%%%%%%%%%%%%%%%%%%%%%%%%%%%%%%%%%%%%%%%%%%%%%%%%%%%%%%

\subsection{Localized basis functions}

\begin{figure}
  \centering
    \subfloat[]{\includegraphics[width=0.45\columnwidth]{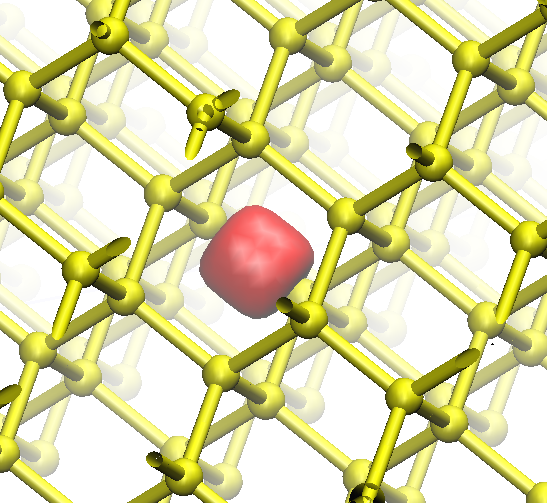}}
    \quad
    \subfloat[]{\includegraphics[width=0.45\columnwidth]{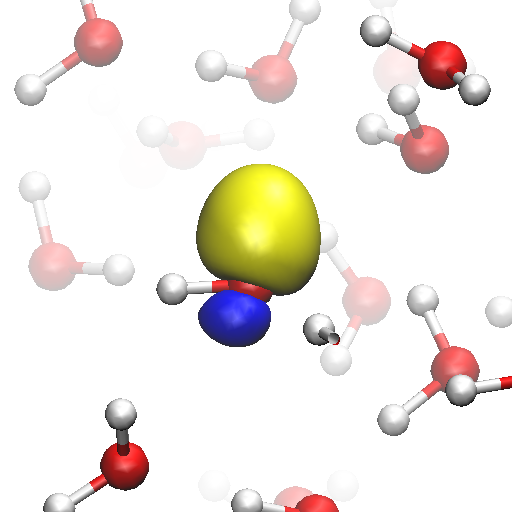}}
    \caption{(color online) Isosurface for an orthogonalized SCDM for: (a) A silicon crystal with
    $512$ Si atoms (yellow balls). The red isosurface characterizes the orthogonalized
    SCDM located between two Si atoms.  (b) A water system with $64$ O atoms (red balls) and
    $128$ H atoms (white balls). The yellow and blue isosurfaces
    characterize the positive and negative part of the orthogonalized SCDM
    respectively. }
    \label{fig:orbitals}  
\end{figure}

We now demonstrate the effectiveness of our algorithm from both
qualitative and quantitative points of view.
For all of our numerical experiments we used {\sc Quantum
ESPRESSO}~\cite{QE} to
compute the eigenfunctions. The SCDM are then computed from the 
Kohn-Sham orbitals for the occupied states. 
Fig.~\ref{fig:orbitals} (a) and (b) show one of the orthogonalized SCDM obtained from a
silicon crystal on a cubic domain with 512 atoms, and a snapshot of a water system with
$64$ molecules.  The kinetic energy cutoff for the silicon
crystal is $10$ Ry, and for water is $75$ Ry. For all
calculations we use the Troullier-Martins
pseudopotential~\cite{TroullierMartins1991}
with the Perdew-Burke-Ernzerhof (PBE)
functional~\cite{PerdewBurkeErnzerhof1996}. 
The orbitals are very localized in the real space and resemble the
shape of MLWFs~\cite{MarzariVanderbilt1997}.  Our method automatically finds the centers
of all localized orbitals, which for the silicon crystal are
in the middle of Si-Si bond, and for water is closer to the oxygen atoms
than to the hydrogen atoms.

\begin{figure}
	\centering
	\subfloat[]{\includegraphics[width=0.45\columnwidth]{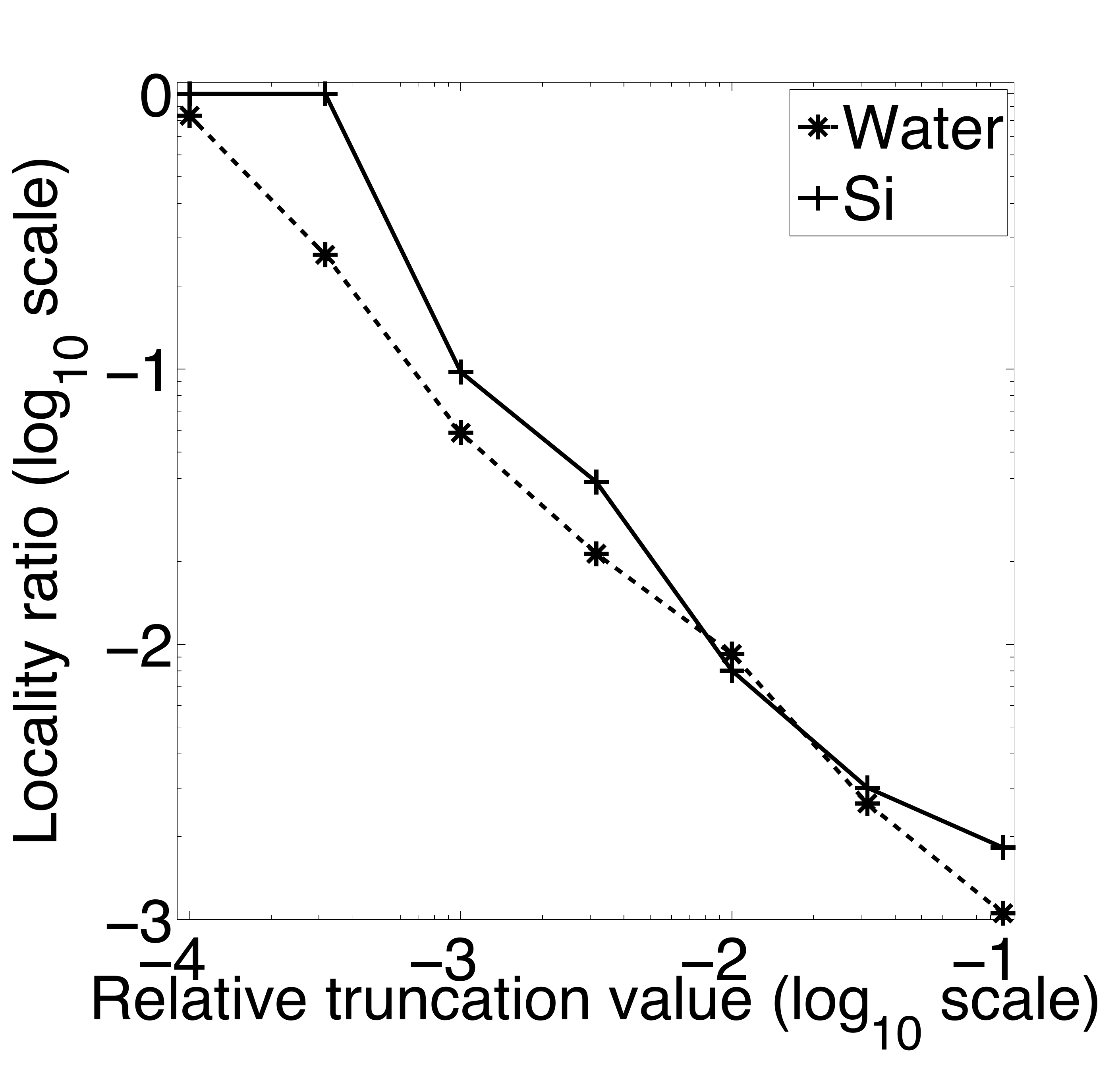}}
	\quad
	\subfloat[]{\includegraphics[width=0.45\columnwidth]{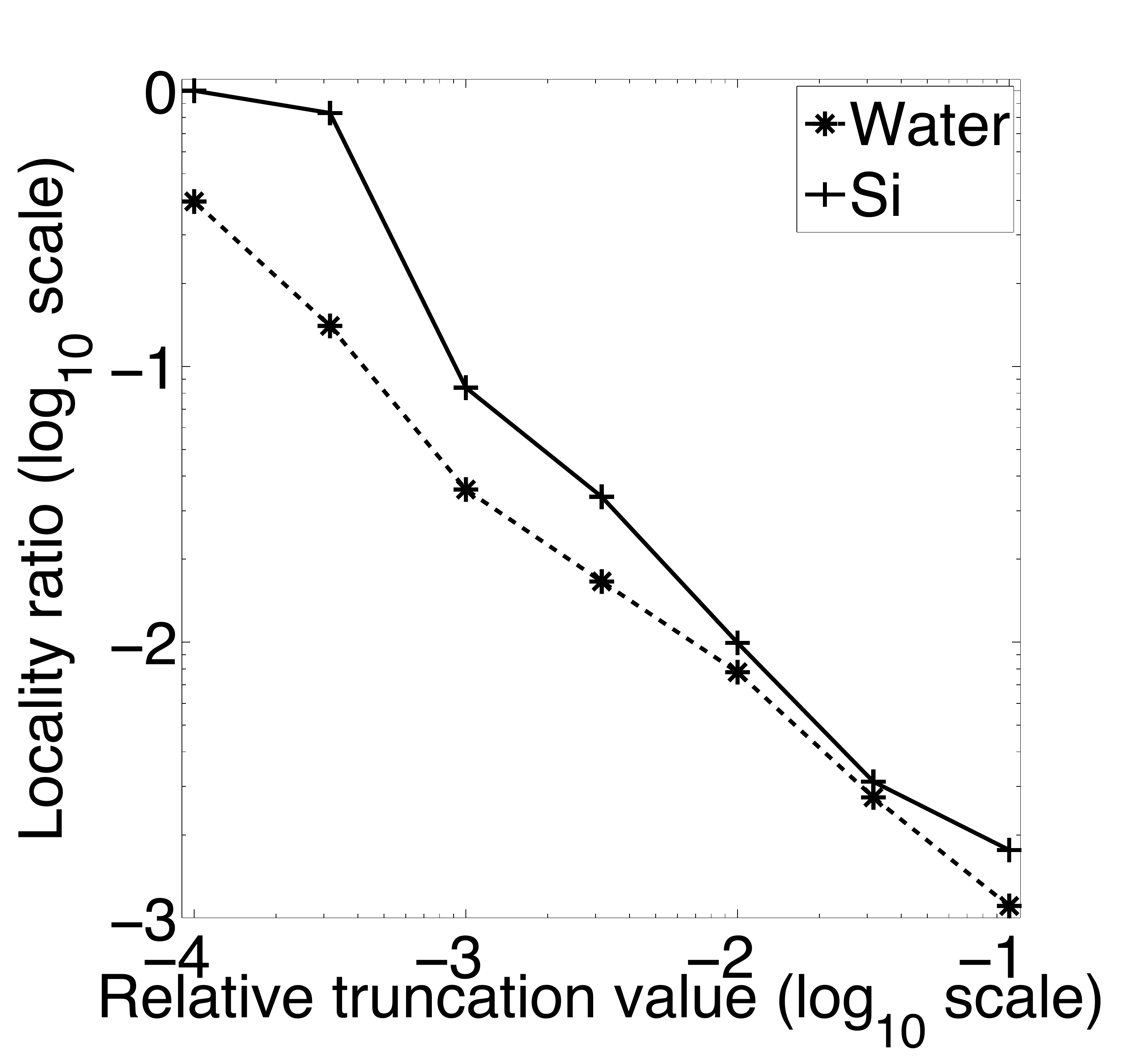}}
    \caption{Average locality of basis functions for silicon crystal and water
    molecules; (a) non orthogonal and (b) orthogonal. (See text for the
    definition of the locality ratio)
    }
    \label{fig:locality}  
\end{figure}

The locality of the SCDM is demonstrated by computing the
locality ratio. We define this as the ratio between the volume of a
cuboid domain that contains the support of the basis function after
truncating the functions below a certain relative threshold and the volume of the whole domain.
Fig.~\ref{fig:locality} shows the average locality ratio of
the SCDM basis functions, both before and after orthogonalization, for each test system as the relative truncation threshold
is varied. We clearly
see the exponential decay of the SCDM for silicon and water.
When the relative truncation value is $10^{-2}$, the support of each
SCDM only occupies $1\sim 2\%$ of the computational domain, therefore
significantly reducing the storage cost for the orbitals. Furthermore, as anticipated the orthogonalization procedure does not impact the locality of the functions, and in some cases even serves to further localize them. 
The good localization properties also implies that $P_{\CS,\CS}$
should be a well conditioned matrix. In fact, we observe
that the condition number is only $3.18$ for the silicon crystal and $2.83$ for
the water molecules.

\subsection{Computation of Hartree-Fock exchange energy}

We now use the orthogonalized SCDM to compute the Hartree-Fock exchange energy with, under some mild assumptions,
linear scaling cost. The Hartree-Fock exchange energy is invariant
under unitary transformation of the Kohn-Sham orbitals, and can be
computed as
\begin{equation}
\label{eqn:hfee}
E_{x} = -\frac{1}{2} \sum_{i,j=1}^{N_e} \iint
\frac{\phi_{i}(x)\phi_{j}(x) \phi_{j}(y) \phi_{i}(y)}{|x-y|} \ud x \ud
y. 
\end{equation}
Here $\{\phi_{i}\}_{i=1}^{N_e}$ can be the Kohn-Sham orbitals, or any
unitary transformation of them.  When Eq.~\eqref{eqn:hfee} is computed
using the Kohn-Sham orbitals $\{\psi_{i}\}_{i=1}^{N_e}$ directly the
standard method is to use the planewave basis set and solve Poisson's
equation for each pair of Kohn-Sham orbitals.   This results in having
to solve $\Or\left(N_e^2\right)$ discrete versions of Poisson's equation
with periodic boundary conditions and enforcing zero mean in the
solution, which may be done via use of the Fast Fourier Transform (FFT).
The overall cost for computing the energy is then $\Or\left(N_e^2 N \log
N\right),$ where the $\log N$ factor arises from the FFT. In a hybrid
functional calculation such as PBE0
functionals~\cite{PerdewErnzerhofBurke1996} the cost of this step can be dominating
compared to a calculation using semi-local exchange-correlation
functionals.

By using the orthogonalized and truncated SCDM instead of the delocalized
Kohn-Sham orbitals the computational cost can be significantly reduced,
at the expense of introducing some controllable error. Let
$\hat{\Phi}=[\hat{\phi}_{1},  \ldots,  \hat{\phi}_{N_e}]$ denote the truncated
version of $\Phi.$ 
This truncation introduces some error, but the relative truncation value may be chosen
to achieve any desired accuracy.
After truncation we may simply neglect pairs of $\hat{\phi}_{i}$ and
$\hat{\phi}_{j}$ with disjoint support, or if their product is
sufficiently small on the global domain. When each $\hat{\phi}_{i}$ is
localized the $\Or\left(N_e^2\right)$ terms in Eq.~\eqref{eqn:hfee} is
reduced to $\Or\left(N_e\right)$ significant terms. The computational cost can be further
reduced by noting that the solution of Poisson's equation is only needed
on the support of $\hat{\phi}_{i} \hat{\phi}_{j}.$ This fact may be used
to reduce the size of each FFT by using Hockney's
algorithm~\cite{Hockney1965,EastwoodBrownrigg1979}.
Hockney's algorithm is a fast and direct method that does not introduce any
additional approximation error.  A brief presentation of the main idea of Hockney's
method is provided in the supporting information.

Given the above techniques we still have to determine the support of
$\hat{\phi}_{i} \hat{\phi}_{j}$ for all $i,j$ pairs, which scales as
$\Or(N_e^2).$ However, if the vectors are appropriately stored this
operation amounts to a few logical operations per pair of $\hat{\phi}$
and will not be a dominant portion of the computational cost. Ignoring
this cost and assuming that as the number of atoms and size of a
molecule grows the support of the basis functions remains constant
yields an overall computational scaling of $\Or\left(N_e\right).$ Here
the constant will dominantly depend on the size of FFTs required by
Hockney's algorithm.

\begin{figure}
	\centering
	\subfloat[]{\includegraphics[width=\columnwidth]{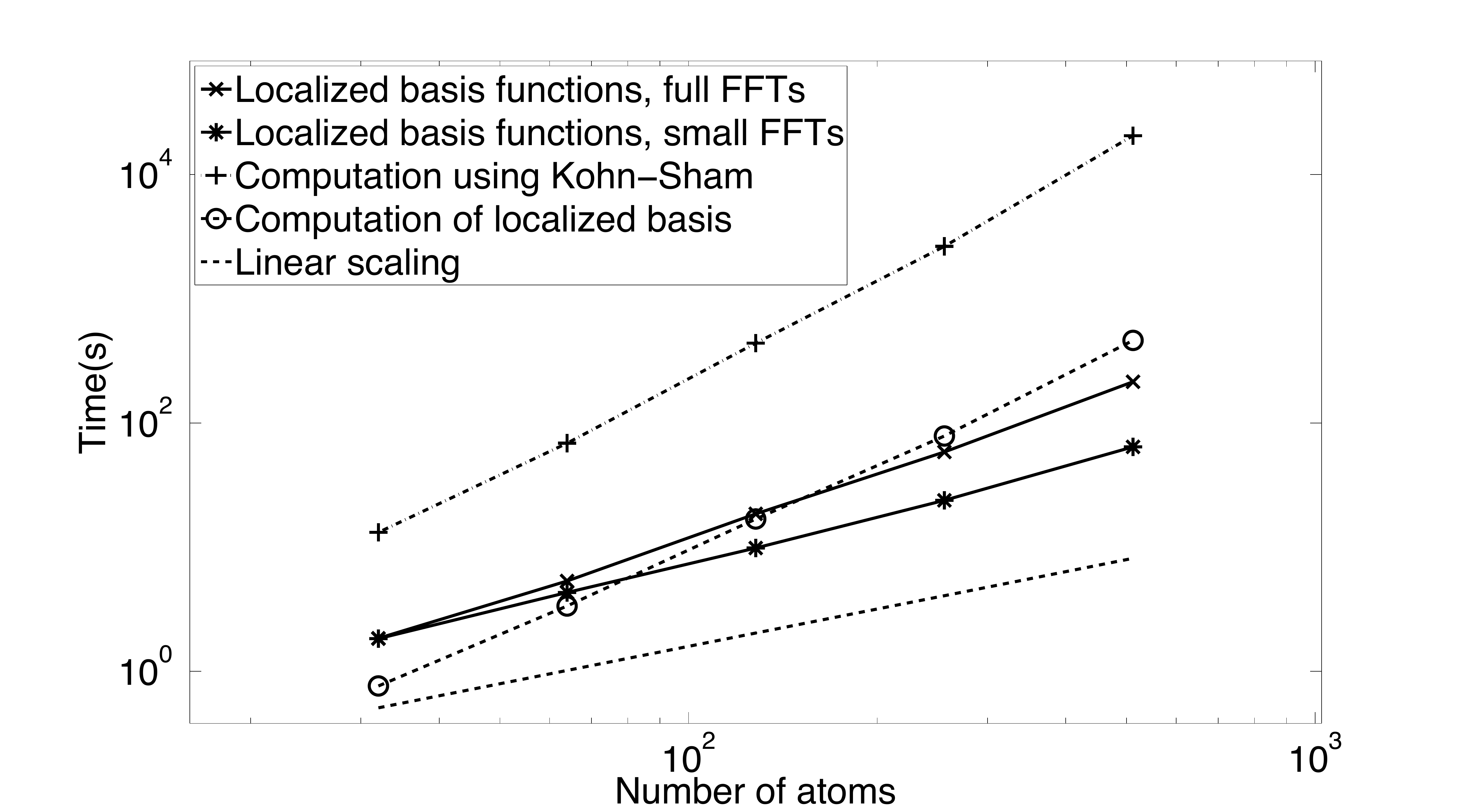}} \\
	\subfloat[]{\includegraphics[width=\columnwidth]{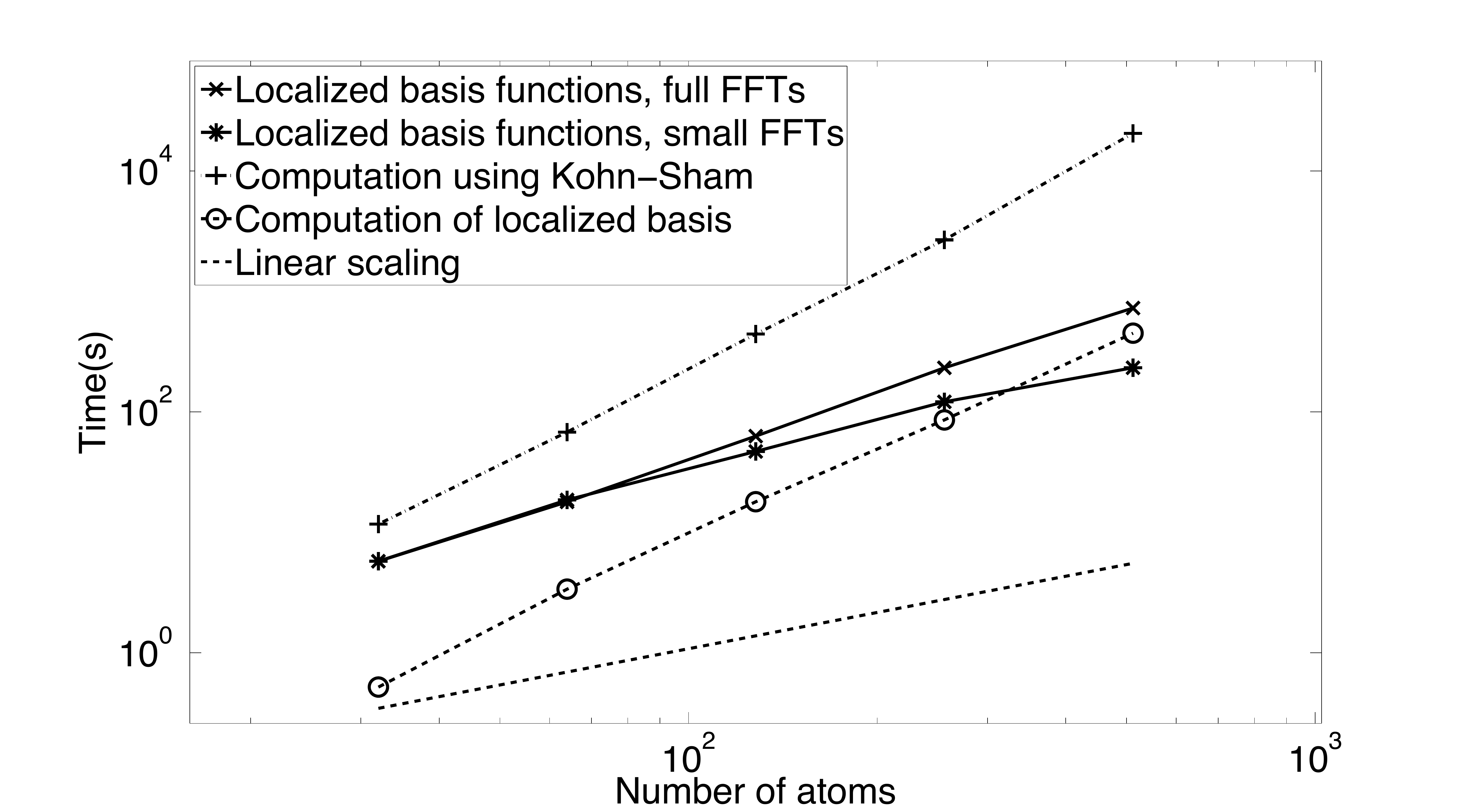}}
    \caption{Time to compute the Hartree-Fock exchange energy and the
    localized basis functions. The localized basis functions were
    truncated to achieve (a) 4\% - 5\%  and (b) 0.3\% - 0.4\% relative
    error in the energy computation.}
    \label{fig:hfee}  
\end{figure}

To demonstrate the efficiency and accuracy of computing the Hartree-Fock
exchange energy using the orthogonalized SCDM we construct a quasi
one-dimensional silicon crystal extended along the $z$-direction. The
total number of atoms varies from 32 to 512 and for each
problem size $N_e$ is twice the number of atoms. The kinetic
energy cutoff and other parameters are the same as for the
aforementioned silicon crystal. All calculations are performed on a
single computational node.   Fig. \ref{fig:hfee} shows the time to compute the
Hartree-Fock exchange energy when using the localized and truncated
basis functions. Our error criteria is the relative error in the
computed exchange energy. We truncate the localized basis functions at
two different values, and achieve relative error of $4\%\sim 5\%$ in
Fig.~\ref{fig:hfee} (a) and $0.3\sim 0.4\%$ in Fig.~\ref{fig:hfee} (b).
We observe that using the localized and truncated functions greatly
reduces the computational time even when factoring in the cost of
performing the localization to obtain the orthogonalized SCDM. The cost
for the localization scales as $\Or(N N_{e}^2)$ but is over an order of
magnitude faster (even more so if a parallel version is used) than computing the exchange energy directly using
Kohn-Sham orbitals.  It is also clear that the use of localized basis
functions without using Hockney's algorithm (``full FFTs'') reduces the
observed computational cost to $\Or(N N_{e} \log{N})$, while using
Hockney's algorithm (``Hockney'') further reduces the computational cost
to $\Or(N_{e}).$ For the largest problem size the speedup in computing
the exchange energy using the localized basis functions and Hockney's
algorithm relative to the computational time using Kohn-Sham orbitals is
about a factor of 300 when the relative error is $4\%\sim
5\%$, and  a factor of 90 when the relative error is $0.3\%\sim
0.4\%.$

\subsection{Parallel computation}

\begin{figure}
  \centering
  \includegraphics[width=\columnwidth]{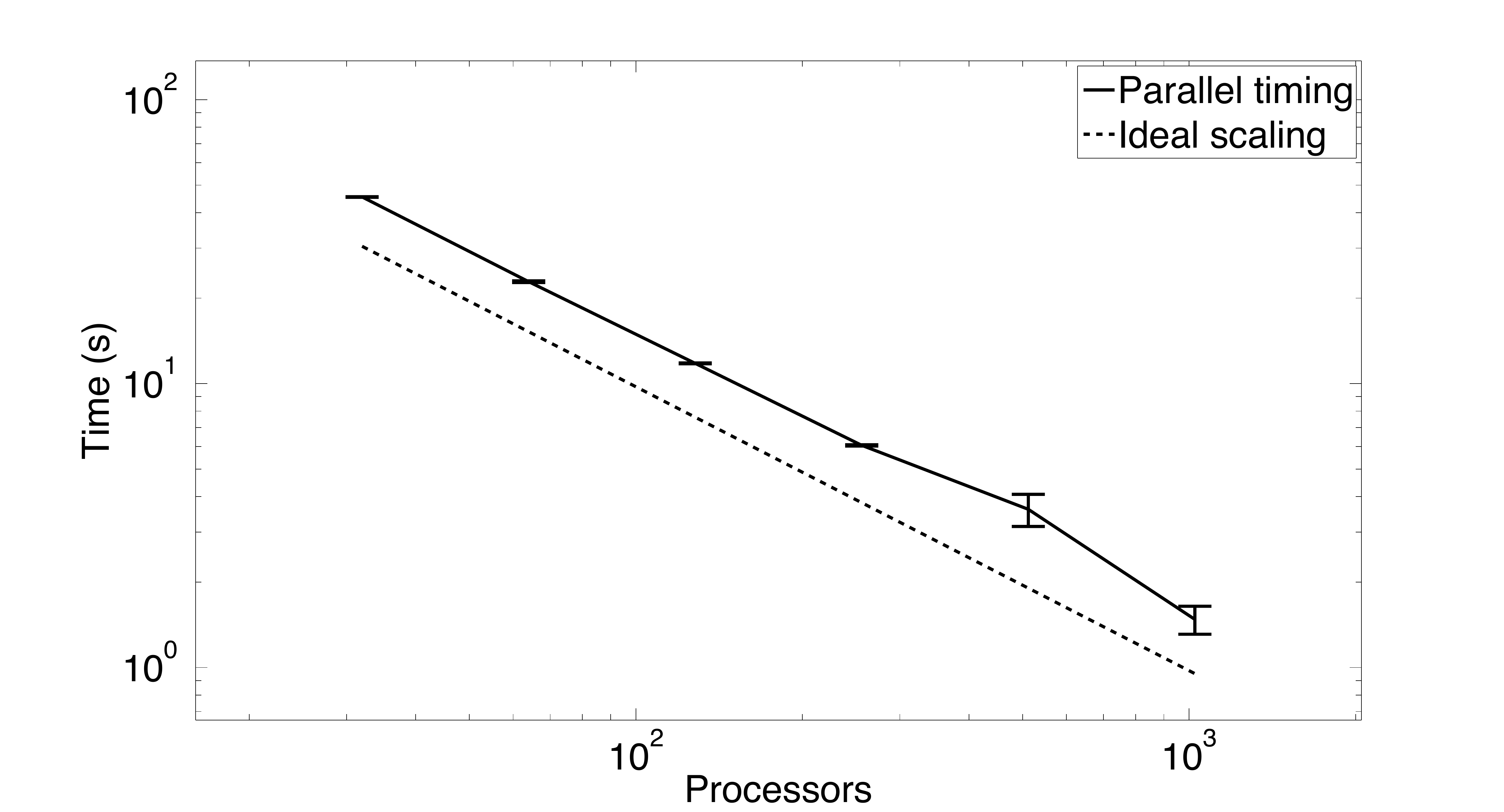} \caption{Parallel
  scaling for computing the orthogonalized SCDM. The times given are the
  average over 10 computations, and the error bars are one standard
  deviation away from the mean.}
  \label{fig:parallel} 
\end{figure}

The above discussion featured results for the localization procedure and
the subsequent computation of the Hartree-Fock exchange energy when
using a single machine. For the largest quasi one-dimensional problem
considered $\Psi$ is a $777600 \times 1024$ matrix. To demonstrate how the algorithm
may be parallelized we implemented a simple parallel version in FORTRAN
using ScaLAPACK \cite{Scalapack} for the parallel rank-revealing QR. The pivoted QR, matrix multiplication
to form $\hat{\Phi},$ and the triangular solve to form $\Phi$ were all
done using parallel versions of the respective algorithms. We make note of the fact that the parallel timings given here are not on machines of the same type as those used previously for the examples, so the absolute times are not directly comparable. Computing the
orthogonalized SCDM for the aforementioned problem using this code and
1024 processors took under 1.5 seconds. This is not including the time
to distribute $\Psi$ amongst the processors, which took 1.3 seconds, nor
the time to read $\Psi$ from disk. In Figure \ref{fig:parallel} we show
the scaling of this parallel implementation for computing the
orthogonalized SCDM. Once again we omit the time taken to distribute the
matrix. Here we see that for the given problem size the method scales
close to ideally on up to 1024 processors.

%%%%%%%%%%%%%%%%%%%%%%%%%%%%%%%%%%%%%%%%%%%%%%%%%%%%%%%%%%%%%%%%%%%%%%%%%%%%%%%
\section{Conclusion}\label{sec:conclusion}
%%%%%%%%%%%%%%%%%%%%%%%%%%%%%%%%%%%%%%%%%%%%%%%%%%%%%%%%%%%%%%%%%%%%%%%%%%%%%%%
We have presented a simple procedure to obtain compressed Kohn-Sham
orbitals by directly using SCDM. The nearsightedness principle
guarantees the locality of the SCDM for insulating systems. The computation of the
orthogonalized SCDM is a simple and fast procedure that may be done
immediately following the computation of $\Psi$ to build a well
localized orthogonal basis for the Kohn-Sham orbital subspace. Because the
algorithm is built out of a few very simple linear algebra routines,
it is simple to implement, parallelize, and include in 
electronic structure software packages. 

Our work can be extended in several directions.  Besides the computation
of the Hartree-Fock exchange energy, the high parallel scalability of
the SCDM procedure can enable the computation of localized basis
functions for the self-consistent treatment of the Hartree-Fock
exchange terms in KSDFT calculations with hybrid functionals.

In this work we
explicitly require the Kohn-Sham orbitals to be represented on a real
space grid.  This is very natural for electronic structure software
packages based on planewave and finite difference methods, but may or
may not be natural for other basis sets such as localized atomic
orbitals.  In fact, the locality of the single particle density matrix
along the off-diagonal direction holds in general when the density
matrix is represented using localized basis
functions~\cite{BenziBoitoRazouk2013}.  Therefore it would be possible
to directly find the localized representation of the single particle
density matrix represented using localized basis functions and skip the
real space representation.  The numerical consequence of this procedure
should be carefully tested. 

The cost of computing the SCDM may be reduced via a randomized version
of the algorithm. For the examples presented here this method slightly
reduced the quality of the localized basis functions and consequently
was not used. However, it may still be beneficial for very large
problems. The SCDM may also be a useful tool for achieving linear
scaling electronic structure calculations for insulating systems, as
well as higher level quantum chemical treatment at the post-DFT level.
Our idea is not limited to the compression of Kohn-Sham orbitals, but
may also be generalized for the compression of pair products of
Kohn-Sham orbitals in excited state calculations.

% \section*{Supporting Information}
% A brief overview of Hockney's algorithm is provided.

%%%%%%%%%%%%%%%%%%%%%%%%%%%%%%%%%%%%%%%%%%%%%%%%%%%%%%%%%%%%%%%%%%%%%%%%%%%%%%%
\section*{Acknowledgments}
%%%%%%%%%%%%%%%%%%%%%%%%%%%%%%%%%%%%%%%%%%%%%%%%%%%%%%%%%%%%%%%%%%%%%%%%%%%%%%%
% \begin{acknowledgements}
This work is partially supported by NSF Fellowship DGE-1147470 (A. D.)
and NSF grant DMS-0846501 (A.D. and L. Y.); by a Simons Graduate Research Assistantship (A. D.); by the DOE Scientific
Discovery through Advanced Computing (SciDAC) program, and the DOE
Center for Applied Mathematics for Energy Research Applications (CAMERA)
program (L. L.); and by the Mathematical Multifaceted Integrated Capability Centers (MMICCs) effort within the Applied Mathematics activity of the U.S. Department of Energy’s Advanced Scientific Computing Research program, under Award Number DE-SC0009409 (L. Y.).  We thank
Lenya Ryzhik and the National Energy Research Scientific Computing
(NERSC) center for providing the computational resources. We are
grateful to Wibe de Jong for his valuable suggestions to improve our manuscript.
% \end{acknowledgements}

\bibliography{localize}

\end{document}